\documentclass[reprint,superscriptaddress,showpacs,nofootinbib,amsmath,amssymb]{revtex4-1}
\usepackage[sort&compress]{natbib}
\usepackage{graphicx}
\usepackage{epstopdf}
\usepackage{subfigure}
\usepackage{float}
\usepackage{dcolumn}
\usepackage{hyperref}
\usepackage{bm}
\usepackage{color}

\begin{document}

\title{Novel ansatz for  charge radii  in density functional theories}
\author{Rong An}
\affiliation{School of Physics,
Beihang University, Beijing 100191, China}

\author{Li-Sheng Geng}
\email[E-mail: ]{lisheng.geng@buaa.edu.cn}
\affiliation{School of Physics, Beihang University, Beijing 100191, China}
\affiliation{
Beijing Key Laboratory of Advanced Nuclear Materials and Physics,
Beihang University, Beijing 100191, China}
\affiliation{Beijing Advanced Innovation Center for Big Data-Based Precision Medicine, School of Medicine and Engineering, Beihang University, Beijing, 100191}
\affiliation{School of Physics and Microelectronics, Zhengzhou University, Zhengzhou, Henan 450001, China}

\author{Shi-Sheng Zhang}
\email[E-mail: ]{zss76@buaa.edu.cn}
\affiliation{School of Physics,
Beihang University, Beijing 100191, China}

\date{\today}

\begin{abstract}
  Charge radii are one of the most fundamental properties of atomic nuclei characterizing their
   charge distributions. Though the general trend as a function of the mass number
   is well described by the $A^{1/3}$ rule, some fine structures, such as the evolution
   along the calcium isotopic chain and the corresponding odd-even staggerings, are
   notoriously difficult to describe both in density functional theories and ab initio methods. In
    this letter, we propose a novel ansatz to describe the charge radii of calcium isotopes,
    by adding a correction term,  proportional to the number of Cooper pairs, and determined by
  the BCS amplitudes and a single parameter, to the charge radii calculated in the
  relativistic mean field model with the pairing interaction treated with the BCS method.
  The new ansatz yields results
  consistent with data not only for calcium isotopes, but also for ten other isotopic chains, including oxygen, neon,
  magnesium, chromium, nickel, germanium, zirconium, cadmium, tin, and lead. It is
  remarkable that this ansatz with a single parameter can  describe nuclear charge radii throughout the
  periodic table, particularly
  the odd-even staggerings and parabolic behavior. We hope that the present study can
  stimulate more discussions about its nature and relation with other effects proposed to explain the odd-even staggerings of charge radii.
\end{abstract}


\maketitle
{\it Introduction:} Charge radii~\cite{Angeli:2013epw}, just as masses~\cite{Wangmeng30002},
are among the most fundamental quantities to characterize the ground states of atomic nuclei.
On the other hand, compared to masses, they seem to be harder to describe theoretically.  For instance,
ab initio calculations have been able to describe masses (or binding energies) of light (or even medium mass) nuclei reasonably well for quite some time (see, e.g., Ref.~\cite{Pieper:2001mp}),
but failed to do so for charge radii until quite recently (see, e.g, Refs.~\cite{Soma:2019bso,Arthuis:2020toz} and references therein). Nowadays, various semi-microscopic mass models~\cite{Wang:2014qqa} or mean field theories~\cite{Geng:2005yu,Goriely:2016uhb} are
able to describe known nuclear binding energies with an accuracy of less than 3 MeV, with a discrepancy of much
less than 1\%. As for charge radii, the discrepancy is still at the level of 5\%~\cite{Geng:2005yu,Goriely:2016uhb}. In particular, there are
a few long-standing discrepancies in the theoretical description of charge radii. One of the most intricate  is
about calcium isotopes, which show a parabolic behavior between~$^{39}$Ca and~$^{48}$Ca and strong odd-even staggerings~\cite{Angeli:2013epw,NP-12-594,nature15}.  Furthermore,
 $^{52}$Ca has a radius larger than that of ~$^{48}$Ca~\cite{NP-12-594},
  though $N=32$~\cite{nature498} and 34~\cite{nature502} are found to be new magic numbers.  All these features
have remained a challenge for better theoretical understanding.


Conventional density functional theories, such as the relativistic mean field (RMF) model (see below) or  the Hartree-Fock-Bogoliubov method (see, e.g., Fig.~6 of Ref.~\cite{Saperstein:2017har}), can not describe the peculiar features of calcium isotopes. The
DF3-a EDF model can describe  the charge radii of
$^{39-49}$Ca, but then fails to do so for $^{50-52}$Ca~\cite{Tolokonnikov2010}.
The extended DF3-a+ph model, in which the particle-phonon
coupling effects  are taken into
account~\cite{Saperstein:2017har}, improves the description of $^{50-52}$Ca.
The Fayans EDF model, with a novel density-gradient term added to the pairing interaction~\cite{FAYANS200049}, can somewhat reproduce the odd-even staggerings below $N=28$~\cite{PhysRevC.95.064328,nature15}, but not so satisfactorily (see Fig.~1).
 Ab initio calculations with chiral EFT interactions, such as $\mathrm{NNLO}_\mathrm{sat}$~\cite{PhysRevC.91.051301},
  SRG1 and SRG2~\cite{PhysRevC.83.031301},
  fail to reproduce experimental data as well~\cite{NP-12-594}.

It is clear that none of the existing theoretical models can describe the charge radii of
calcium isotopes satisfactorily, which implies that some important physics is missing
in all these models.  In the present work, we propose a novel  ansatz based on the
 RMF model and we show that it can describe the charge radii of calcium isotopes rather well. Such an ansatz is
then extended to study ten other isotopic chains, and is shown to work remarkably well. How can such a term is derived in a more microscopic way
in the RMF model is not clear yet, but its origin is very clear, i.e., the neutron proton pairing,
 as the evolution of charge radii along an isotopic chain can only come from the interaction between neutrons
and protons. The odd-even staggering then clearly shows that this effect is from the pairing channel.

{\it Novel ansatz for charge radii:} The RMF model within non-linear Lagrangian densities
is used in this study~(see Refs.~\cite{Vretenar:2005zz,Meng:2005jv,
jie2016relativistic} for some latest reviews). The corresponding Dirac equation for
nucleons and Klein-Gordon equations  for mesons and the photon
 are solved by the expansion method with the harmonic
oscillator basis~\cite{Ring:1997tc,Geng:2003pk}. In the present
work, 12 shells are used to expand the fermion fields and 20 shells for
the meson fields. The mean-field effective force used is NL3~\cite{PhysRevC.55.540}.~\footnote{Using other effective forces
such as TM1~\cite{SUGAHARA1994557} and PK1~\cite{PhysRevC.69.034319} does not
change essentially any of our conclusions, but does affect the description of charge radii at a
quantitative level.}
In order to describe the pairing correlation, the state-dependent BCS
method with a delta force is employed~\cite{Geng:2003pk}, with its strength tuned for each isotopic chain.
The odd-A nuclei are treated with the blocking approximation.
A detailed description of the deformed RMF(BCS) method can be found in Refs.~\cite{Geng:2003pk}.

In the RMF(BCS) method, conventionally charge radii are calculated in the following way~\cite{Ring:1997tc,Geng:2003pk}:
\begin{eqnarray}
R_{ch}^{2}&=&\frac{\int{r}^{2}d^{3}n_{p}(r)}{\int{d}^{3}n_{p}(r)}+0.64~\mathrm{fm^{2}},
\end{eqnarray}
where the first term represents the charge distribution of  point-like protons and the second
term accounts for the finite size of the proton.

 As can be seen from Fig.~1, the RMF(BCS) method cannot describe the charge radii of calcium isotopes,
 particularly, the parabolic behavior between $N=20$ and $N=28$ and the odd-even staggerings. We have checked that
 neither the point coupling version of the RMF model, or the deformed relativistic Hartree Bogoliubov theory in continuum~\cite{Li:2012gv}, can describe calcium isotopes. As discussed in the introduction, the Skyme Hartree-Fock methods also fail to describe calcium isotopes. This is also the case for ab initio methods. See Refs.~\cite{NP-12-594,nature15} for more discussions.

A closer inspection of the experimental data reveals the following: the charge radius of $^{40}$Ca and that
of $^{48}$Ca are almost the same, which can be understood due to the closed shell at $N=20$ and $N=28$. In between, the charge radii are larger, but more interestingly, show odd-even staggerings, similar to those more familiar in the binding energies of atomic nuclei.  All these strongly point to the fact that such features are related to the pairing interaction. As the proton number is fixed,  the evolution of the charge radius along the calcium isotopic chain should be attributed to neutron-proton correlations.
Inspired by the above observation, we propose to add a correction term to the charge radius of Eq.~(1), such that it becomes:
\begin{eqnarray}\label{coop1}
R_{ch}^{2}=\frac{\int{r}^{2}d^{3}n_{p}(r)}{\int{d}^{3}n_{p}(r)}+0.64~\mathrm{fm^{2}}+\frac{a_{0}}{\sqrt{A}}\cdot\Delta{\mathcal{D}}~\mathrm{fm^{2}}
\end{eqnarray}
In the above expression, $A$ is the mass number, the constant
 $a_{0}=0.834$ is a normalization constant fixed by fitting to
 the charge radius of $^{44}$Ca, and the quantity
 $\Delta\mathcal{D}=|\mathcal{D}_{n}-\mathcal{D}_{p}|$ is defined as
 \begin{eqnarray}
 \mathcal{D}_{n,p}=\sum^{n,p}_{k>0}u_{k}v_{k},
 \end{eqnarray}
 where  $v_k$ and $u_k$ are the BCS amplitudes, with $v_k^2$ the occupation
 probability of single particle orbital $k$, $u_k^2=1-v_k^2$, and the summation is over
 all the occupied single particle levels.
   It should be emphasized in the present work that all the quantities appearing in the
  correction term are obtained self-consistently in the RMF(BCS) method.

 It is interesting to note that $u_{k}v_{k}$ represents a measure of the number of correlated pairs (Cooper pairs)~\footnote{The BCS coherent pairs mix correlated and
 uncorrelated pairs over the whole system, and therefore they cannot be interpreted as
 Cooper pairs except in the extreme strong-coupling
 and dilute limits where all the pairs are bounded and non-overlapping~\cite{PhysRevA.72.043611}} in the BCS wave function~\cite{cooperpair}
\begin{eqnarray}
u_{k}v_{k} =\langle\mathrm{BCS}|a_{k}^{\dagger}a_{\bar{k}}^{\dagger}|\mathrm{BCS}\rangle.
 \end{eqnarray}
 Therefore, $\Delta\mathcal{D}$ represents the difference of the fractions of Cooper
 pairs between neutrons and protons. In a recent work, Miller et al. has shown that in
 ab initio calculations,  charge radii might be influenced by short range correlations missing in soft nucleon-nucleon interactions ~\cite{Miller:2018mfb}.
 Our proposed ansatz [Eq.~(2)] somewhat resembles Eq.~(9) of Ref.~\cite{Miller:2018mfb}.

\begin{figure}[htbp]
    \includegraphics[scale=0.4]{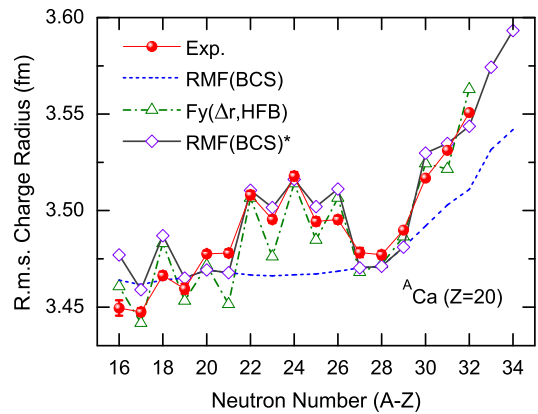}
    \caption{Charge radii of calcium isotopes obtained by the RMF(BCS) method and our new ansatz, denoted by RMF(BCS)*. The experimental data are taken from Refs.~\cite{Angeli:2013epw,NP-12-594,nature15}. For comparison the Fayans EDF results~\cite{nature15} are also shown. \label{fig1}}
\end{figure}

{\it Results and discussions:} In Fig.~\ref{fig1}, we compare the charge radii of  calcium isotopes calculated in the
RMF(BCS) method with and without the correction term. It is clear that
without the correction term, the RMF(BCS) results cannot describe experimental data, particularly, the parabolic behavior, the odd-even staggerings, as well as the fast increase beyond $N=28$. On the other hand, with the correction term, the RMF(BCS)* results describe
the data quite well.  In fact, the results are even better than those of the Fayans EDF model, particularly for those nuclei with $N>20$~\cite{nature15}.  Although the RMF(BCS)* method overestimates the charge radii of $^{16,17,18}$Ca, it does
correctly yield  the odd-even staggerings (see also the lower panel of Fig.~2).
It will be interesting to see how future experimental data compare with our predictions for calcium isotopes with $N\ge53$.
\begin{figure}[htbp]
   \includegraphics[scale=0.34]{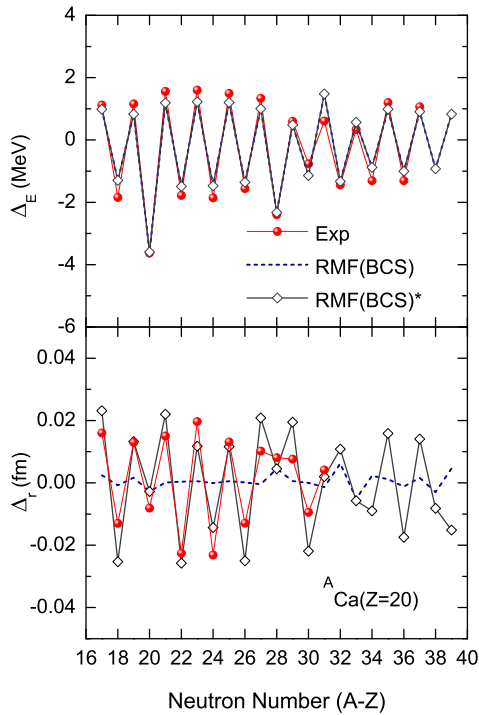}
    \caption{Odd-even staggerings in the binding energies (upper panel) and charge radii (lower panel) of calcium isotopes. The experimental data of binding energies are taken from Ref.~\cite{Wangmeng30002}, while those of charge radii are from Refs.~\cite{Angeli:2013epw,NP-12-594,nature15}. \label{fig2}}
\end{figure}

To highlight effects of the pairing correlation (and eliminating a smooth background from the mean field part), for bulk properties, such as  masses and charge radii, one can
define the so-called ``double odd-even staggerings'' for either
an isotopic chain or an isotonic chain. For binding energies,
one can use the following three-point formula~\cite{P.Ring,A.Bohr}:
\begin{eqnarray}\label{oes}
\Delta_E&=&\frac{1}{2}[B(N-1,Z)-2B(N,Z)+B(N+1,Z)],\\\nonumber
\end{eqnarray}
where the $B(N,Z)$ is the binding energy for a nucleus of neutron number
$N$ and proton number $Z$. Similarly, one can define a
 three-point formula to extract odd-even staggerings for charge radii~\cite{PhysRevC.95.064328}:
\begin{eqnarray}
\Delta_r&=&\frac{1}{2}[R(N-1,Z)-2R(N,Z)+R(N+1,Z)],
\end{eqnarray}
where  $R(N,Z)$ is the root-mean-square charge radius.

In Fig.\ref{fig2}, the odd-even staggerings of nuclear
masses (upper panel) and charge radii (lower panel) are compared with
experimental data. It is clear the RMF(BCS)/RMF(BCS)*\footnote{They give the same results for
 masses.} method reproduces the odd-even staggerings of the masses
quite well (as expected), but fails miserably for charge radii, while
the new ansatz is able to describe  charge radii remarkably well.
We note that around the
magic number N$=28$, the odd-even staggering effects seem to slightly overestimated.

The new ansatz can describe the charge radii of calcium isotopes
very well, implying that it must have captured some important physics. Naturally, one would like to
see whether it works for other nuclei. For such a purpose, we study the
 charge radii of oxygen, nickel, germanium, and zirconium, whose proton number is either magic or semi-magic, and neon, magnesium, chromium, and cadmium isotopes.

\begin{figure}[htbp]
\includegraphics[scale=0.45]{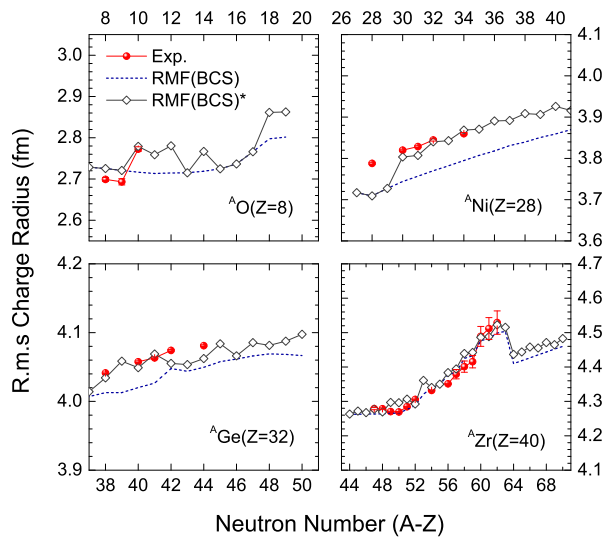}
 \caption{Charge radii of oxygen, nickel, germanium, and zirconium isotopes obtained by the RMF(BCS) method and our new ansatz RMF(BCS)*. The experimental data are taken from Refs.~\cite{Angeli:2013epw,NADJAKOV1994133}. \label{fig3}}
\end{figure}

In Fig.~\ref{fig3}, the charge radii of oxygen, nickel, germanium, and zirconium isotopes
are compared with experimental data. For oxygen isotopes,
the new ansatz yields results in agreement with the experimental data,
especially, the sharp increase from $^{17}$O to $^{18}$O.
For nickel isotopes, the new ansatz can reproduce the experimental
data except for $^{56}$Ni, with  $N=28$ a magic number (see discussions below). For germanium isotopes, the charge radii are also better reproduced  by the new ansatz, notably those of $N=38,40,41$. While for zirconium isotopes, the new ansatz barely change the results, in agreement with data.

\begin{figure}[htbp]
\includegraphics[scale=0.45]{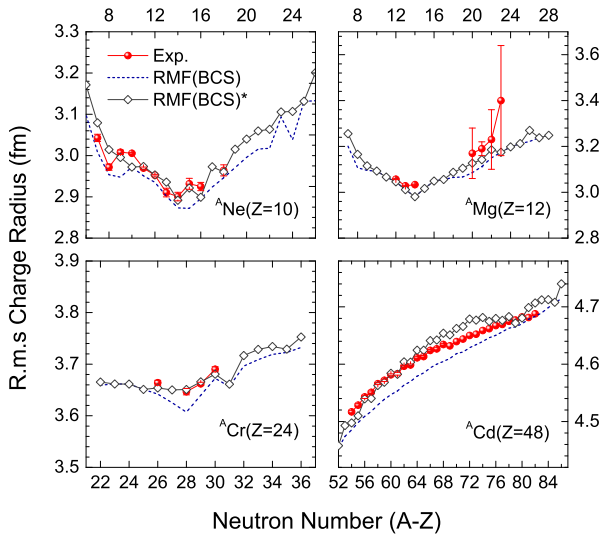}
 \caption{Charge radii of neon, magnesium, chromium, and cadmium isotopes obtained  by the RMF(BCS) method and our new ansatz RMF(BCS)*. The experimental data are taken from Refs.~\cite{Angeli:2013epw,NADJAKOV1994133,PhysRevLett.121.102501}. \label{fig4}}
\end{figure}

In Fig.~\ref{fig4}, the charge radii of  neon, magnesium, chromium, and cadmium isotopes are shown. The charge radii of neon
 isotopes are better described by the new ansatz, especially the odd-even staggerings of  $^{24-26}$Ne. For magnesium isotopes,  the new ansatz barely changes the results of the RMF(BCS) method.
For chromium isotopes, the results from the new ansatz are in better agreement with the
experimental data, particularly for $^{52}$Cr.
For most cadmium isotopes, the new ansatz again yields better results. Only around $A=120$, the charge radii are overestimated. One should note that the normalization factor of the correction term is
fixed by $^{44}$Ca.
It is interesting to point out that here once more we see a clear parabolic behavior between $N=50/64$ and $N=82$, similar to  calcium isotopes

\begin{figure}[htbp]
\includegraphics[scale=0.45]{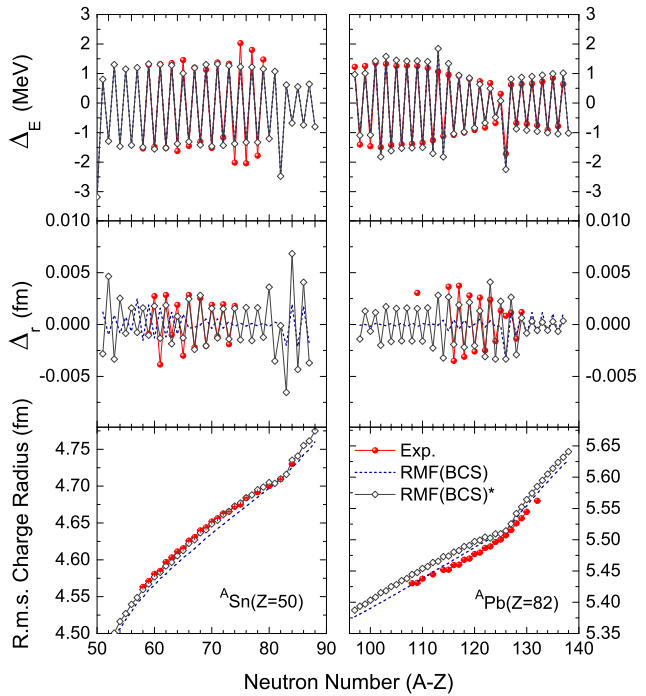}
   \caption{Odd-even staggerings in the binding energies (upper) and charge radii (middle) and the charge radii (bottom) of tin and lead isotopes
   obtained by the RMF(BCS) method and our new ansatz RMF(BCS)*. The experimental data of binding energies are taken  from Ref.~\cite{Wangmeng30002} and those of charge radii are from Ref.~\cite{Angeli:2013epw,NADJAKOV1994133,Gorges:2019wzy}. \label{figdoe}}
\end{figure}

Next, we investigate whether the new ansatz works for heavy nuclei with $Z\ge50$. More specifically, we study
tin and lead isotopes, whose $\Delta_E$, $\Delta_r$, and charge radii are shown in Fig.~\ref{figdoe}. The theoretical results
 agree well with the experimental data for both  binding energies and charge radii. It should be noted that for tin and lead isotopes
 we have to use a smaller $a_0=0.22$ fixed by the charge radius of $^{126}$Sn to better produce the odd-even staggerings in
 the charge radii~\footnote{It should be noted that for lead isotopes, the RMF(BCS) results already
 overestimate the charge radii but underestimated the corresponding odd-even staggerings.}. This reflects that the strength of the correction term might need to be adjusted for different isotopes. Or in other words, its mass dependence is not completely captured in the $1/\sqrt{A}$ factor.~\footnote{This is understandable because
  even for the pairing strength of the delta force one has to re-tune it a little bit for different isotopic chains to better reproduce the odd-even staggerings of binding energies.} In addition, we note that the discontinuities across $N=82$~\cite{Gorges:2019wzy} and
  $N=126$~\cite{Goddard:2012dk} are well reproduced in the RMF method (which has been attributed to the rather small
  isospin dependence of the spin-orbit term in the RMF method~\cite{Sharma:1994mim}).
A peculiar observation regarding lead isotopes is that the theoretical odd-even staggerings of charge radii around $N=126$
is larger than its experimental counterpart, though they are smaller than data for those nuclei away from $N=126$. It seems for nuclei with either magic proton or neutron numbers,
the new ansatz behaves relatively worse. This is true also for those calcium isotopes nearby  $N=28$. Such a feature might be
related to non-conservation of particle number in the BCS method.

{\it Summary and outlook:} Motivated by the peculiar features exhibited in the charge radii of calcium isotopes and the fact
mean-field calculations failed to describe most (if not all) of the fine structures, we proposed a novel  ansatz to
correct the charge radii obtained from the RMF(BCS) method. With a single parameter fixed by
the charge radius of $^{44}$Ca, we were able to reproduce the charge radii of calcium isotopes from $^{40}$Ca up to $^{54}$Ca,
in a way comparable or even better than most refined existing calculations. This ansatz was then applied to study ten more isotopic chains, i.e., oxygen, neon,
  magnesium, chromium, nickel, germanium, zirconium, cadmium, tin, and lead. Overall remarkable improvements are achieved, though for tin
  and lead isotopes we had to use a smaller normalization constant.  In particular, we noted that the parabolic
behavior of cadmium isotopes is also reasonably reproduced.

The more microscopic origin of the new  ansatz remains to be identified. The most plausible factor is the
 short-range neutron-proton pairing correlation, which is missing
in (most) mean-field calculations. Regardless of its microscopic nature, the phenomenological successes demonstrated in the present work hints that it must have captured  relevant physics. In the future, one may further test such an ansatz by studying all the isotopic chains for which experimental data exist. In doing so, slight readjustment of the mean-field parameters might be needed as well. Furthermore, as the ansatz is based on the BCS theory, where particle number is not conserved, one may wish to study how its restoration affects the description of charge radii using, e.g., the FBCS method~\cite{RongAn:114101}. The fact that our ansatz performed slightly worse for nuclei with magic numbers, such as
 $N=28$ and $N=126$, indeed indicates such an necessity. Last but not the least, the new ansatz is also expected to work for
  Skyme Hartree-Fock models and  therefore explicit studies are strongly encouraged.

{\it Acknowledgments:} This work is supported in part by the National
 Natural Science Foundation of China under Grants Nos.11735003, 11975041, 11775014, and 11961141004

\bibliography{refs}

\end{document}